\documentclass[preprint]{revtex4}
\usepackage{amsfonts}
\usepackage{amsmath,color}
\usepackage{graphicx}

\newcommand{\ea}{{\it et al.}}

\begin{document}

\title{Recommended isolated-line profile for representing high-resolution spectroscopic transitions (IUPAC Technical Report)
}
\author{Jonathan Tennyson}
\email{j.tennyson@ucl.ac.uk}
\affiliation{Department of Physics and Astronomy, University College London,
Gower Street, London WC1E 6BT, UK}
\author{Peter F. Bernath}
\affiliation{Department of Chemistry and Biochemistry,
Old Dominion University,
Norfolk, VA 23529 USA}
\author{Alain Campargue}
\affiliation{Universit\'e Grenoble 1/CNRS, UMR5588 LIPhy, Grenoble, F-38041 France}
\author{Attila G. Cs\'asz\'ar}
\affiliation{Institute  of Chemistry, Lor\'and E\"otv\"os University,
H-1117 Budapest, P\'azm\'any s\'et\'{a}ny 1/A, Hungary and MTA-ELTE Research Group on Complex
Chemical Systems, H-1518 Budapest 112, P.O. Box 32, Hungary}
\author{Ludovic Daumont}
\affiliation{GSMA, UMR CNRS 7331, Universit\'e de Reims Champagne Ardenne
U.F.R. Sciences Exactes et Naturelles
Moulin de la Housse B.P. 1039
51687 Reims Cedex 2, France}
\author{Robert R. Gamache}
\affiliation{Department of Environmental, Earth, and Atmospheric Sciences
University of Massachusetts Lowell
Lowell, MA 01854, USA}
\author{Joseph T. Hodges}
\affiliation{National Institute of Standards and Technology, 
Gaithersburg, MD, U.S.A.}
\author{Daniel Lisak}
\affiliation{Institute of Physics, Faculty of Physics, Astronomy and Informatics, Nicolaus Copernicus University, Grudziadzka 5, 87-100 Torun, Poland}
\author{Olga V. Naumenko}
\affiliation{Institute of Atmospheric Optics, 
Russian Academy of Sciences, Tomsk, Russia}
\author{Laurence S. Rothman}
\affiliation{ Atomic and Molecular Physics Division,
Harvard-Smithsonian Center for Astrophysics, Cambridge, MA, U.S.A.}
\author{Ha Tran}    
\affiliation{Laboratoire Interuniversitaire des Syst\`{e}mes Atmosph\'{e}riques, UMR CNRS 7583, Universit\'{e} Paris Est Cr\'{e}teil, Universit\'{e} Paris Diderot, 
Institut Pierre-Simon Laplace, 94010 Cr\'{e}teil Cedex, France}
\author{Nikolai F. Zobov}
\affiliation{Institute of Applied Physics, Russian Academy of Sciences, 46 Uljanov
Street, 603950 Nizhny Novgorod, Russia}
\author{Jeanna Buldyreva}
\affiliation{Institut UTINAM UMR CNRS 6213, Universit\'e de Franche-Comt\'e, 16 route de Gray, 25030 Besan\c{c}on, France}
\author{Chris D. Boone}
\affiliation{Department of Chemistry, University of Waterloo, 200 University Ave W, Waterloo, Ontario, N2L 3G1, Canada}
\author{Maria Domenica De Vizia and Livio Gianfrani}
\affiliation{Department of Mathematics and Physics, Second University of Naples, Caserta, Italy}
\author{Jean-Michel Hartmann}
\affiliation{Laboratoire Interuniversitaire des Systemes Atmosph\'{e}riques, UMR CNRS 7583, Universit\'{e} Paris Est Cr\'{e}teil, Universit\'{e} Paris Diderot, 
Institut Pierre-Simon Laplace, 94010 Cr\'{e}teil Cedex, France}
\author{Robert McPheat}
\affiliation{Space Science and Technology Department,
Rutherford Appleton Laboratory,
Harwell Oxford Campus,
Didcot, OX11 0QX, UK}
\author{Jonathan Murray}
\affiliation{Space and Atmospheric Physics,
Imperial College London, London, Prince Consort Road,
London SW7 2BW, UK}
\author{Ngoc Hoa Ngo}
\affiliation{Faculty of Physics, Hanoi National University of Education, 136 Xuan Thuy, Cau Giay, Hanoi, Vietnam}
\author{Oleg L. Polyansky}
\affiliation{Department of Physics and Astronomy, University College London,
Gower Street, London WC1E 6BT, UK\\ and Institute of Applied Physics, Russian Academy of Sciences, 46 Uljanov
Street, 603950 Nizhny Novgorod, Russia}
\author{Damien Weidmann}
\affiliation{Space Science and Technology Department,
Rutherford Appleton Laboratory,
Harwell Oxford Campus,
Didcot, OX11 0QX, UK}

\date{\today}

\begin{abstract}
The report of an IUPAC Task Group, formed in 2011 on
``Intensities and line shapes in high-resolution spectra of water
isotopologues from experiment and theory'' (Project No.
2011-022-2-100), on line profiles of isolated high-resolution
rotational-vibrational transitions perturbed by neutral gas-phase 
molecules is presented. 
The well-documented inadequacies of the Voigt profile (VP), used almost
  universally by databases and radiative-transfer codes, to represent
  pressure effects and Doppler broadening in isolated
  vibrational-rotational and pure rotational  transitions of
the water molecule have resulted in the development of a variety of alternative
  line-profile models. 
These models capture more of the physics of
the influence of pressure on line shapes but, in
  general, at the price of greater complexity.  
The Task Group recommends that the partially Correlated quadratic-Speed-Dependent
  Hard-Collision profile should be adopted as the appropriate model for
  high-resolution spectroscopy.
For simplicity this should be called the Hartmann--Tran profile (HTP).
The HTP  is sophisticated enough to capture the various
  collisional contributions to the isolated line shape, can be
  computed in a straightforward and rapid manner, and reduces to
  simpler profiles, including the Voigt profile, under certain
  simplifying assumptions.
 
\end{abstract}


\maketitle

\section{Introduction}

Characterization of an isolated spectral line measured under high resolution
requires three pieces of information: the transition frequency, the transition integrated intensity, and the parameters that describe the line profile.
A previous IUPAC Task Group (hereafter called TG1), 
comprising several of the current authors, has
critically evaluated the  line frequency data available
for all the major isotopologues of water  \cite{jt454,jt482,jt539,jt576}.
A summary of this work along with recommendations of TG1 were 
recently presented in this journal \cite{jt562}. 

The full characterization of the high-resolution spectrum of water vapour from the
microwave to the ultraviolet is a prerequisite for modelling and
understanding of various processes in many fields in chemistry, physics, earth sciences and engineering.
The diverse areas of interest include:
\begin{enumerate}
\item Atmospheric modelling,
with emphasis on the definitive understanding of global warming as water
vapour is responsible for about
70 \%\ of the known absorption of sunlight and 
the majority of the greenhouse effect;
\item Atmospheric remote sensing and environmental monitoring,
since it is generally necessary to remove the spectral signature of water
in order to interpret correctly the signatures from
trace species; 
\item Satellite communication,
as the performance of satellites in the Earth's atmosphere is sensitive to
water absorption between about 3 and 400 GHz;
\item Active remote sensing such as radar and lidar that is affected by water vapour attenuation;
\item Studies of planetary and exoplanetary atmospheres;
\item Astronomy, for example, that of
cool stars, where hot water is a major constituent; 
water lasers and masers, which are widespread in outer space, and the study of 
comets based on fluorescence spectroscopy;
\item Combustion research, 
such as rocket exhausts, forest fires, and turbine engines, 
as hot steam is a major product of most combustion processes.
\end{enumerate}

One of the recommendations of TG1 was the urgent need
to identify and adopt a reference line profile for high-resolution
spectroscopic studies which improved
upon the current standard, the so-called Voigt profile (VP). 
The present paper reports the related recommendation
of another IUPAC Task Group (hereafter called TG2) on
``Intensities and line shapes in high-resolution spectra
of water isotopologues from experiment and theory''
(Project No. 2011-022-2-100). TG2 limited itself to considering gas-phase transitions which occur
in neutral environments as the pressure effects in plasmas need a somewhat
different treatment.

The line profile of an isolated spectroscopic transition is usually 
defined as being normalized to unit area and can be
attributed
to the following three physical factors:
\begin{enumerate}
\item The Heisenberg time-energy uncertainty principle, 
or, equivalently, the spontaneous emission of radiation, 
is responsible for the natural lifetime broadening or \textit{intrinsic line width}. 
This component of the overall line shape is described by a Lorentzian profile 
which is, however, sufficiently narrow 
to be safely neglected in favour of the next two contributions in all but the most specialized situations
and ultra-low temperatures;
\item The thermal translational motion of the spectroscopically active molecule at velocity $v_a$ gives 
the incident radiation, of frequency $\nu_0$, a frequency shift of $\Delta\nu$ = $\pm(v_a/c)\nu_0$ 
in the molecular frame of reference: the well-known \textit{Doppler effect}. 
The corresponding Doppler profile (DP) is expressed in terms of the Doppler half-width, $\mathit{\Gamma}_{\rm D}$, by a Gaussian function:
\begin{equation}
F_{\rm D}(\nu - \nu_0) = \sqrt{{\ln(2) \over \pi}}{1 \over \mathit{\Gamma}_{\rm D}}\exp{(-\ln(2)({{\nu - \nu_0}\over {\mathit{\Gamma}_{\rm D}}}})^2).
\end{equation} 
For temperature $T$, in K, and molecular mass $m$, in kg, the Doppler half-width, in Hz, is
\begin{equation}
\mathit{\Gamma}_{\rm D} =\sqrt{2\ln(2)kT \over {mc^2}}\nu_0=1.459 313 6 (7)\times10^{-20}\sqrt{\rm {(kg/K)}}\sqrt{(T/m)}\nu_0,
\end{equation}
or equivalently in terms of the molar mass, $m_\textrm{m}$ in g mol$^{-1}$, 
$\mathit{\Gamma}_{\rm D}=3.581 163  (2)\times10^{-7}\sqrt{\rm {g\over mol} {1\over K}} \sqrt{(T/m_{\textrm{m}})}\nu_0$.

\item Individual collisions of molecules lead to energy exchanges between radiators and perturbers. 
These exchanges shorten the lifetime  of the initial and final states of the optical transition 
and yield what is called pressure or \textit{collisional broadening}. 
These collisions  also induce  pressure-dependent shifts in the central frequency 
of the transition. Assuming the independence of the pressure-broadened line half-width at 
half-maximum, $\mathit{\Gamma}$, and of the 
pressure-induced line shift, $\mathit{\Delta}$, from the molecular speeds (mean thermal velocity approximation), 
one obtains, for the associated profile, a homogeneous Lorentzian function:
\begin{equation}
F_{\rm L}(\nu-\nu_0)=\frac{1}{\pi} \frac{\mathit{\Gamma}}{(\nu-\nu_0-\mathit{\Delta})^2+\mathit{\Gamma}^2}.
\end{equation}
\end{enumerate}

At low pressures the Doppler effect dominates, and as the pressure
increases the effects of collisions become increasingly important.  As
a first approximation to get the resulting line shape, the convolution
of an inhomogeneous Doppler profile with a homogeneous Lorentzian
profile is commonly used.  It defines the so-called VP,
which contains Doppler and Lorentzian shapes as limiting cases.  
The three parameters, $\mathit{\Gamma}_{\rm D}$, $\mathit{\Gamma}$ and $\mathit{\Delta}$,
characterizing the Voigt profile are routinely employed in standard
spectroscopic information systems \cite{jt557,jt504}.  
$\mathit{\Gamma}_{\rm D}$ is independent of the gas mixture composition and has a known temperature dependence, see eq. 2. 
For $\mathit{\Gamma}$ and $\mathit{\Delta}$, their values in a mixture are simply obtained, assuming binary collisions, 
through model fraction-weighted average of the individual values for each collision partner. The temperature 
dependence of $\mathit{\Gamma}$ and $\mathit{\Delta}$ is commonly assumed to be a power law for the broadening \cite{jt557,jt504} but 
remains to be found for the pressure shift.
Although the
Voigt profile involves an integral that cannot be evaluated
analytically, there are readily available, fast computational
procedures for doing this, see, {\it e.g.}, \cite{79Hum,07KeBe}, which make
this function suitable for use in complex radiative-transfer codes.

There is now a widespread recognition that the VP does not
give a fully accurate representation of the spectral line shape
\cite{08HaBoRo} and its use can lead, for example, to a systematic
underestimation of experimental line intensities
\cite{08LiHo,12Kochanov,12NgIbLa}.  Use of the VP for
modelling spectra of water vapour recorded in both the laboratory and
the atmosphere under high resolution leads to characteristic W-shaped
residuals in any high-precision fit to the line absorption
coefficient.  See, for example,
Refs.~\cite{01ClHeHu,06LiHoCi,07TrBeDo,11DeRoCa} for laboratory work
and Refs.~\cite{05DuZePa,07BoWaBe} for atmospheric studies.  In
particular, large W-shaped residuals were observed in the analysis of
H$_2$O lines from the Atmospheric Chemistry
Experiment~\cite{07BoWaBe}, an effect that appears to occur only for
H$_2$O and not for other molecules in the spectra.  Deviations from
the Voigt line shape contribute to the large residuals for H$_2$O
lines, although there are likely other contributions relating to the
high variability of the H$_2$O column in the Earth's
atmosphere and the rapid change of H$_2$O
volume mixing ratio as a function of altitude in the troposphere,
particularly in the tropics.

The W-shaped residuals arise, and their amplitude may reach 10~\%\ of
the peak absorption \cite{07TrBeDo,12NgTrGa}, as the observed lines
are typically higher and narrower than predicted by the VP.
Deviations from the VP are generally ascribed to the effect
of velocity changes (VCs) due to collisions, which reduce $\mathit{\Gamma}_{\rm D}$, 
and/or to the speed dependence (SD) of the relaxation rates,
which corrects the simple Lorentzian shape for different
velocity-classes of active molecules.  Inclusion of these velocity
effects in line-profile models has led to the development of a variety
of possible line shape functions which can be characterized by their
increasingly sophisticated representation of the underlying physics
and an increasing number of parameters. Several of these models are
briefly considered in Section III.

\section{Objectives of the Task Group}
There are many suggested models for the line profile which move beyond
the VP
\cite{Galatry,NG,Rautian,72EnCaHaKe,72Berman,73EnKeMoCa,79EnMa}; see
chapter III of the book by Hartmann \ea\ \cite{08HaBoRo} for a more
general discussion.  These profiles are more sophisticated in that
they include more physical effects than are accounted for by the VP
and, as a consequence, require additional model parameters.  At
present, databases and most major radiative-transfer codes use Voigt
profiles, despite their well-documented deficiencies.  The
implementation of profiles beyond the Voigt approximation in databases
and codes represents a major task and will only be undertaken once
there is some reasonable consensus as to which line profile to use.
This has become particularly timely as increased sensitivity of remote
sounding instruments and better knowledge of geophysical parameters
has increased the relative significance of line-profile issues for
observing systems.  The aim of TG2 and of this article is therefore to
recommend the use of a single functional form suitable for line
profiles representing high-resolution spectroscopic transitions.

To achieve this goal the following points need to be considered:
\begin{enumerate}
\item What single functional form from those available is most 
appropriate to replace the VP? 
The chosen function should
have a sound theoretical basis and behave in an appropriate manner
as a function of pressure and temperature.
\item Is the chosen functional form computationally tractable?
Radiative-transfer models which perform line-by-line
calculations for large numbers of lines demand a functional form that
can be evaluated reliably and efficiently.
\item What are the consequences of replacing Voigt functions in databases and models? 
There should be a straightforward path from the current situation which relies on Voigt 
functions to the newly recommended function.
\end{enumerate}
To answer these questions, the
water molecule serves as a suitable benchmark. 
As detailed in the Introduction, water line profiles are of great importance for a number of applications.
Furthermore, water has a large permanent dipole moment which leads to
strong long-range interactions, making  the line shape of its
transitions particularly challenging to model from the purely theoretical
point of view. 
TG2 therefore aimed to make
a recommendation for water which would also be appropriate and 
adopted for other molecules.

Finally, we note that collision-induced mixing of nearby lines also alters line
profiles, see chapter IV of Ref.~\cite{08HaBoRo} for
a general discussion of this process. 
This effect has been observed in water transitions
\cite{bbd05,12PeSoSo} but  was not explicitly considered by TG2. 
Nevertheless, the significance of the TG recommendations for 
the inclusion of line mixing
and for the water continuum is discussed below.

\begin{table}
\caption{Summary of line-profile models considered. $N$ is the
number of parameters required to characterize the line shape
for a single isolated transition at a given temperature for
a given pair of molecules. See text for further details and citations.
All profiles except the simple Lorentz profile include the Doppler broadening effect.  \\
}
 \label{tab:profiles}
\tiny
\begin{tabular}{llclclll}
\hline \hline
Acronym & Profile name& \multicolumn{2}{l}{Parameters} &&\multicolumn{3}{l}{Mechanism}\\ \cline{3-4}\cline{6-8}
        &    & $N$ &            &&SD$^a$ & VC$^a$ & Correlation \\
\hline
DP      & Doppler  & 1  & $\mathit{\Gamma}_{\rm D}$ &&No & No & No\\
LP      & Lorentz  & 2  & $\mathit{\Gamma}$, $\mathit{\Delta}$ && No & No & No\\
VP      & Voigt & 3 & $\mathit{\Gamma}_{\rm D}$, $\mathit{\Gamma}$, $\mathit{\Delta}$ &&  No & No & No\\
GP      & Galatry & 4& $\mathit{\Gamma}_{\rm D}$, $\mathit{\Gamma}$, $\mathit{\Delta}$, $\nu_{\rm VC}$ && No & Soft & No\\
RP      & Rautian & 4& $\mathit{\Gamma}_{\rm D}$, $\mathit{\Gamma}$, $\mathit{\Delta}$, $\nu_{\rm VC}$ && No& Hard & No\\
NGP     & Nelkin--Ghatak & 4&  $\mathit{\Gamma}_{\rm D}$, $\mathit{\Gamma}$, $\mathit{\Delta}$, $\nu_{\rm VC}$ && No& Hard & No\\
SDVP$^b$    & speed-dependent Voigt & 5 & $\mathit{\Gamma}_{\rm D}$, $\mathit{\Gamma}_0$, $\mathit{\Delta}_0$,  $\mathit{\Gamma}_2$, $\mathit{\Delta}_2$&& Yes & No & No\\
SDGP$^b$    &speed-dependent Galatry & 6& $\mathit{\Gamma}_{\rm D}$, $\mathit{\Gamma}_0$, $\mathit{\Delta}_0$,  $\mathit{\Gamma}_2$, $\mathit{\Delta}_2$, $\nu_{\rm VC}$  && Yes & Soft & No\\
SDNGP$^b$   & speed-dependent Nelkin--Ghatak& 6 &$\mathit{\Gamma}_{\rm D}$, $\mathit{\Gamma}_0$, $\mathit{\Delta}_0$, $\mathit{\Gamma}_2$, $\mathit{\Delta}_2$,  $\nu_{\rm VC}$&& Yes & Hard & No\\
SDRP$^b$    & speed-dependent Rautian& 6 &$\mathit{\Gamma}_{\rm D}$, $\mathit{\Gamma}_0$, $\mathit{\Delta}_0$,   $\mathit{\Gamma}_2$, $\mathit{\Delta}_2$, $\nu_{\rm VC}$&& Yes & Hard & No\\
HTP     & Hartmann--Tran & 7& $\mathit{\Gamma}_{\rm D}$, $\mathit{\Gamma}_0$, $\mathit{\Delta}_0$, $\mathit{\Gamma}_2$, $\mathit{\Delta}_2$, $\nu_{\rm VC}$, $\eta$&&Yes & Hard & Yes\\
CSDaRSP$^b$ & correlated SD asymmetric Rautian--Sobelman &
8 & $\mathit{\Gamma}_{\rm D}$, $\mathit{\Gamma}_0$, $\mathit{\Delta}_0$, $\mathit{\Gamma}_2$, $\mathit{\Delta}_2$, $\nu_{\rm VC}$, $\chi$, $\eta$&& Yes & Combination & Yes\\
pCSDKS$^b$  & partially correlated SD Keilson-Storer &
8 & $\mathit{\Gamma}_{\rm D}$, $\mathit{\Gamma}_0$, $\mathit{\Delta}_0$, $\mathit{\Gamma}_2$, $\mathit{\Delta}_2$, $\nu_{\rm VC}$, $\gamma_{\rm KS}$, $\eta$&& Yes & Combination & Yes\\
\hline \hline
\end{tabular}

$^a$ SD =  speed--dependent;
VC =  velocity changes due to collisions.\\
$^b$ Parameters for these profiles are all given in the quadratic (q) form of the speed dependence; 
for hypergeometric models
the expansion parameters $\mathit{\Gamma}_0$ and $\mathit{\Gamma}_2$ (or $\mathit{\Delta}_0$ and $\mathit{\Delta}_2$)
are replaced by an amplitude factor and a parameter that is either
$p$, the power-law exponent giving the dependence of the broadening on the relative speed,
or $q$, which  describes the power-law dependence of
the intermolecular potential on the intermolecular distance.
\end{table}

\section{Line-profile models: a brief review}
The theory of molecular line shapes is rather complicated and has been
studied over many years \cite{1895Mi}. 
It is not possible to review it all here,
and instead the reader is pointed to two recent books on the topic
\cite{08HaBoRo,10BuLaSt}. 

Atoms and molecules obey the laws of quantum
mechanics and {\it ab initio} quantum mechanical treatments of water
line broadening are available for low-temperature
collisions \cite{10WiFa}, but more approximate treatments are required at higher
temperatures \cite{jt544}. 
This is because at atmospheric temperatures
 a molecule such as water has
too many relaxation channels for a fully quantal treatment of the problem to be practicable. 
The discussion below will therefore be limited to
semi-classical studies. 

When considering the various models, it is also important to remember
that as the models become more sophisticated they have more free
parameters which almost automatically results in better fits of
measured spectra.  However, a high-quality fit to a set of spectral
lines is not a guarantee of the validity of the line-shape model used;
in particular, in the binary-collision regime the fitted parameters
have to be linear functions of the pressure
\cite{06NgBuCo,06LiHoCi,07RoNgBu,07TrBeDo,13BuMaMoRo}.  At the same
time, multispectral fits \cite{95BeRiMaSm,01JaMaDa} of lines recorded
at several pressures of the perturbing gas are particularly useful for
testing and characterizing complicated line-profile functions.
Compared to spectrum-by-spectrum adjustments, multispectrum fits have
the great advantage that they reduce correlations between model
parameters, decrease their uncertainties and make convergence easier.
  Table~\ref{tab:profiles} lists some
of the key line-profile models developed and orders them in terms of
the number of parameters required to characterize a single spectral
transition at a given temperature for a given absorber and a given
perturbing gas.  The standard three-parameter VP, as
already mentioned above, is the simplest line shape accounting for the
pressure and Doppler effects.

The effect of collision-induced velocity changes on spectral line
shape is usually known as Dicke narrowing \cite{Dicke}.  In this case
the strength of the collisions, {\it i.e.},  their efficiency at changing the
velocity, becomes important. Hard-collision models assume that
molecular velocities before and after each collision are completely
decorrelated, {\it i.e.}, each collision is so violent that the molecule
loses completely the memory of its previous velocity and its new
velocity simply follows a Maxwell distribution.  The corresponding
line profile is referred to as a Rautian profile (RP) \cite{Rautian}
or equivalently the Nelkin--Ghatak profile (NGP) \cite{NG}.  The
hypothesis of soft collisions, in which many collisions are necessary
to change the molecular velocity significantly, leads to the Galatry
profile (GP) \cite{Galatry}. Both hard- and soft-collision models
introduce one extra parameter, $\nu_{\rm VC}$, to quantify the
frequency of VC-collisions.

The speed-dependence of the relaxation rates, considered as the single
source of line narrowing, leads to the speed-dependent Voigt profile
(SDVP) \cite{72Berman,Pickett}.  
It should be noted that this speed-dependence can be introduced in more than one way.  
The most popular choice considers a simplified $r^{-q}$ long-range interaction
potential, $r$ being the intermolecular distance and $q=3, 4, 5$, etc.,
for leading dipole--dipole, dipole--quadrupole,
quadrupole--quadrupole, etc., interactions, respectively.  
This approximation results
in an absolute speed-dependence expressed analytically using a confluent
hypergeometric function (often denoted by \lq\lq h" in the model acronym) \cite{74WaCoSm}. 
However, the cost associated with
this model is too high for routine computations, and thus it is common
practice to use a much simpler, quadratic form (given
by a \lq\lq q" in the acronym) \cite{74WaCoSm,94Pine,94RoMaNi,97RoElKaMa}. In this case
the pressure-broadening width and shift are given by
\begin{equation*}
\mathit{\Gamma}(v_{\rm a})=\mathit{\Gamma}_0+\mathit{\Gamma}_2 \lbrack (v_{\rm a}/v_{\rm a0})^2-3/2 \rbrack , 
\end{equation*}
\begin{equation}
\label{eqn_qSD}
\mathit{\Delta}(v_{\rm a})=\mathit{\Delta}_0+\mathit{\Delta}_2 \lbrack (v_{\rm a}/v_{\rm a0})^2-3/2 \rbrack , 
\end{equation}
where $\mathit{\Gamma}_0$ and $\mathit{\Delta}_0$
are, respectively, the collisional width and shift averaged over all speeds,
and the phenomenological
rate parameters $\mathit{\Gamma}_2$ and $\mathit{\Delta}_2$ characterize the dependence on the active-molecule
speed $v_{\rm a}$ ($v_{\rm a0}$ is its most probable value). 
The Boone--Walker--Bernath algorithm \cite{07BoWaBe} allows
the calculation of a SDVP from two VPs.
 
Ascribing the line narrowing solely to the VC-collisions frequently
leads to aberrant values of the $\nu_{\rm VC}$ parameter which can
show unrealistic non-linearities as a function of pressure
\cite{06LiHoCi,07TrBeDo}.  
For example, when just the linear part of
this pressure-dependence is used with a GP to deduce the narrowing
coefficient (the slope), the latter demonstrates higher values than
those allowed by the kinetic diffusion \cite{07RoNgBu,13BuMaMoRo};
conversely, a similar analysis using a RP leads to smaller values of
$\nu_{\rm VC}$ \cite{06LiHoCi}.  
This behavior means that other
treatments of narrowing are needed; in particular, SD should be
accounted for.  To this end, speed-dependence has been introduced in
the soft-collision model, leading to the SDGP \cite{SDGP}.  
The SDGP reduces to the GP in the absence of the speed-dependence ($\mathit{\Gamma}_2=0$
for qSD) and to the SDVP in the absence of velocity-changing
collisions ($\nu_{\rm VC}=0$). Similarly, the SD introduced in the RP
provides the speed-dependent Rautian profile (SDRP) \cite{97LaBlWa}.
For both the SDGP and SDRP there are functional advantages if the
quadratic SD, see eq.~\ref{eqn_qSD}, is assumed
\cite{00PrRoCo,13NgLiTr}.

Both the SDGP and SDRP models assume that the velocity-changing and the
rotational-state-changing aspects of a collision are independent. In
practice this is not true: a change of the  velocities is
balanced by a change of the internal states of the colliders, according to
the energy conservation law.  Therefore, VC and SD mechanisms can operate
simultaneously and their respective model parameters are correlated.
The profile models accounting for the
correlation of these two collisional effects are described by 
functions involving supplementary fitting parameters.
To achieve this correlation requires the introduction of further
parameters. 
An example is the
speed-dependent dispersive Rautian--Galatry profile (SDDRGP) \cite{01PiCi}, which
is used to explain the line-shape asymmetries due to correlation,
hardness and collision duration. The
SDDRGP is a non-analytic line-shape model with many
parameters, which have to be adjusted simultaneously, and
therefore this treatment requires multispectrum fits.

An alternative method of introducing correlation is via the
partially correlated hard-collision model for velocity- and
state-changing collisions \cite{94Pine,99JoBoRo,99Pine}. 
Such a model, based on the use of hypergeometric SD, was recently employed for a
spectroscopic determination of the Boltzmann constant \cite{13MoCaFa}.
However, this hypergeometric form is difficult to use.  A related, but
easier-to-apply model considers the speed-dependence only
quadratically, yielding the partially Correlated
quadratic-Speed-Dependent Hard-Collision Profile (pCqSDHCP)
\cite{13NgLiTr}.  
This model is flexible and has the major advantage
that it can be represented using a relatively simple form, which
allows rapid computational evaluation \cite{13TrNgHa}, an essential
prerequisite for the adoption of a model by databases and modellers.
This model is discussed in detail in the following section.

There are more sophisticated profiles which allow for intermediate
strength collisions, of which the Keilson--Storer (KS) \cite{KS} and
the Rautian--Sobelman (RS) \cite{Rautian} profiles are the most widely
used. These more sophisticated profiles have an additional parameter
$\eta$, the correlation parameter. In the KS model there is also a
memory parameter, $\gamma_{\rm KS}$, which goes to zero for no memory
(a hard collision) and unity for full memory (a soft collision)
\cite{KS,98RoBo}. The RS model instead uses the hardness parameter
$\chi$ \cite{Rautian}.

Of course, the choice of an appropriate
line shape function is not a purely theoretical
excercise and must be guided by fits to high accuracy
measurements, which also need to consider  the appropriate instrumental line
shape function.

\section{The line-profile model}

The line profile recommended by the TG is variously described as the
partially Correlated quadratic-Speed-Dependent Hard-Collision Profile
(pCqSDHCP) or the partially Correlated quadratic-Speed-Dependent
Nelkin--Ghatak Profile (pCqSDNGP).  This line-shape model has been
considered by a number of authors
\cite{94Pine,99JoBoRo,99Pine,01PiCi,13NgLiTr,13TrNgHa}, and has been
used successfully for the  analysis of ultra-high accuracy
experimental water line shapes \cite{06LiHoCi,04LiRuSa,09LiHaHo},
although not all of these studies considered the speed dependence in
quadratic form.  The quoted acronyms represent an attempt to capture
the physics behind the profile but they are hard to remember and
convey little meaning to the non-specialist.  The TG therefore
recommends that this profile, and its computational implementation
which we describe below, be called the Hartmann--Tran profile (HTP)
after the authors of Refs.~\cite{13NgLiTr,13TrNgHa}.

In terms of the 7 parameters 
$\mathit{\Gamma}_{\rm D}$, $\mathit{\Gamma}_0$, $\mathit{\Delta}_0$, 
$\mathit{\Gamma}_2$, $\mathit{\Delta}_2$, $\nu_{\rm VC}$ and $\eta$,
the HTP functional form can be expressed as \cite{13NgLiTr,13TrNgHa,14TrNgHa,14NgLiTr}:
\begin{equation}
F_{\rm HTP}(\nu)=\frac{1}{\pi} {\rm Re}\left\{\frac{A(\nu)}{1-[\nu_{\rm VC}- 
\eta (C_0-3C_2/2)]A(\nu) + (\frac{\eta C_2}{v_{\rm a0}^2})B(\nu)}\right\} .
\end{equation}
The terms $A(\nu)$ and  $B(\nu)$ can be expressed as combinations
of the complex probability function 
\begin{equation}
w(z) = \frac{i}{\pi} \int_{-\infty}^{+\infty} \frac{e^{-t^2}}{z-t} dt =
e^{-z^2} {\rm erfc}(-iz),
\label{eq:w}\end{equation}
where erfc is the Gauss error function, while
\begin{equation*}
 A(\nu) = \frac{\sqrt{\pi}c}{\nu_0 v_{\rm a0}} [w(iZ_-) - w(iZ_+)],
\end{equation*}
\begin{equation}
 B(\nu) = \frac{v_{\rm a0}^2}{\tilde{C}_2}\left[-1 + \frac{\sqrt{\pi}}{2\sqrt{Y}}(1-Z_-^2)w(iZ_-)
- \frac{\sqrt{\pi}}{2\sqrt{Y}}(1-Z_+^2)w(iZ_+)\right].
\end{equation}
In these expressions
\begin{equation*}
 Z_\pm = \sqrt{X+Y} \pm \sqrt{Y},
\end{equation*}
\begin{equation}
X = \frac{-i(\nu_0-\nu)+\tilde{C}_0}{\tilde{C}_2},\ \ \ Y = \left(\frac{\nu_0v_{\rm a0}}{2c\tilde{C}_2}\right)^2,
\end{equation}
where 
\begin{equation*}
 \tilde{C}_0 = (1 - \eta)(C_0 - \frac{3C_2}{2}) + \nu_{\rm VC},
\end{equation*}
\begin{equation}
  \tilde{C}_2 = (1 - \eta)C_2,
\end{equation}
with $C_n = \mathit{\Gamma}_n + i \mathit{\Delta}_n$ with $n=0$ and 2 within the
quadratic approximation, see eq.~\ref{eqn_qSD}.  
Finally, the most probable speed can be expressed in terms of the Doppler half-width as
$v_{\rm a0}= \frac{c}{\sqrt{\ln 2}\nu_0}\mathit{\Gamma}_{\rm D}$, where $c$ is
the speed of light in vacuum.  Note that the HTP is normalized to unit
area and is generally asymmetric, even if only slightly, due to the
correlation and the speed dependence of the line shift. Furthermore,
in the far wing, when $|\nu - \nu_0|$ is much larger than all other
terms, the HTP reduces to a Lorentzian of half-width $\mathit{\Gamma}_0$.

\begin{table}
\caption{Correspondence between various lower-order models and the limits
of the Hartmann--Tran profile (HTP) \cite{13NgLiTr}.}
 \label{tab:limits}
\small
\begin{tabular}{llll}
\hline \hline
Acronym & Profile& Parameters & Limit of HTP\\ 
        &    &             & \\
\hline
DP      & Gaussian  & $\mathit{\Gamma}_{\rm D}$ & $\mathit{\Gamma}_0=\mathit{\Gamma}_2=\mathit{\Delta}_0=
\mathit{\Delta}_2 = \nu_{\rm VC}=\eta =0$\\
VP      & Voigt &  $\mathit{\Gamma}_{\rm D}$, $\mathit{\Gamma}_0$, $\mathit{\Delta}_0$ & $\mathit{\Gamma}_2=
\mathit{\Delta}_2 = \nu_{\rm VC}=\eta =0$\\
RP      & Rautian & $\mathit{\Gamma}_{\rm D}$, $\mathit{\Gamma}_0$, $\mathit{\Delta}_0$  $\nu_{\rm VC}$ & $\mathit{\Gamma}_2=\mathit{\Delta}_2 =\eta =0$\\
qSDVP & speed-dependent Voigt$^a$ & $\mathit{\Gamma}_{\rm D}$, $\mathit{\Gamma}_0$, $\mathit{\Delta}_0$,  $\mathit{\Gamma}_2$, $\mathit{\Delta}_2$ & $\nu_{\rm VC}=\eta =0$\\
qSDRP & speed-dependent Rautian$^a$ & $\mathit{\Gamma}_{\rm D}$, $\mathit{\Gamma}_0$, $\mathit{\Delta}_0$,  $\mathit{\Gamma}_2$, $\mathit{\Delta}_2$, $\nu_{\rm VC}$  & $\eta =0$\\
\hline \hline
\end{tabular}
\vskip 0.1cm
$^a$ Using the quadratic approximation of eq.~\ref{eqn_qSD} for the speed dependence.
\end{table}

The physics underlying the 7-parameter HTP profile is as follows.  
The speed-dependences of the relaxation rates are represented by four
parameters: two, ($\mathit{\Gamma}_0$,$\mathit{\Gamma}_2$), to represent the line
broadening and two, ($\mathit{\Delta}_0$, $\mathit{\Delta}_2$), to represent the line shift.  
$\mathit{\Gamma}_0$ and $\mathit{\Delta}_0$ are the mean relaxation rates,
while $\mathit{\Gamma}_2$ and $\mathit{\Delta}_2$ are the quadratic terms describing the
speed dependence of relaxation \cite{94RoMaNi,97RoElKaMa}, see eq.~\ref{eqn_qSD}.  
Velocity changes are represented within the
Nelkin--Ghatak hard collision model, which requires a single parameter, $\nu_{\rm VC}$.  
Finally, parameter $\eta$ represents the
partial correlation between velocity and rotational state changes due
to collisions.  The assumptions of quadratic speed dependences and
hard collisions are approximations but, as illustrated below, some of
the errors introduced by these approximations are actually compensated
for by the inherent flexibility of the parameter fits.  This model has
been extensively tested both for water transitions and for those of
other molecules \cite{13NgLiTr} using accurate calculated spectral
shapes \cite{12NgTrGa,09TrHaCh,13HaTrNg,13HaSiBo}.

As discussed in the literature \cite{13NgLiTr},
there are a number of advantages the HTP model offers. 

The first is
that $F_{\rm HTP}$ can be expressed in terms of two (complex) Voigt (or
complex probability) functions, see $w(z)$ as given in eq.~\ref{eq:w}.  
Tran {\it et al.} \cite{13TrNgHa} 
provide a routine for evaluating the HTP built on the Voigt
routine of  Humlicek \cite{79Hum}. 
Numerical tests have shown that the relative accuracy 
of this routine is always better than $10^{-4}$ and that the computer time
requirement is at most only five times that required for the 
computation of a VP.

Secondly, in the case where not all the parameters have been or can be
determined, HTP reduces in a very straightforward fashion
to a number of simpler, standard profiles. 
These limiting cases are listed in Table~\ref{tab:limits}. 
In particular, given that the present databases
are largely populated with parameters for the VP, 
it is advantageous that the HTP
reduces to the VP if all the high-order correction terms are
set to zero. 
This is an extremely useful property, but one note of
caution is in order. 
The parameters used to determine these profiles
are not independent; this means that once beyond-Voigt parameters are
introduced into the fit it is no longer possible to use the parameters
to give a correct VP.

Third, we should mention the Van Vleck--Weisskopf (VVW) line shape
function \cite{vVW}, see page 184 of Bernath \cite{95Bernath} for a
short discussion.  This simple form includes an anti-resonant
Lorentzian function.  It is used for low frequency microwave work
\cite{ 00KrTrPa,05TrKoDo,13SlSlGi} and in all models of mm-submm
radiation propagation \cite{89Liebe}.  By analogy with the VVW
profile, the anti-resonant contribution, which is only significant at
very long wavelengths, could be taken into account using an HTP after
changing the signs of the transition frequency and the pressure
shift.

Finally, line-mixing can be easily included in the model provided that
two approximations are made \cite{13NgLiTr}.  The first is the use of
the so-called (Rosenkranz) first-order approximation \cite{75Rosen}.
The second is the neglect of the speed-dependence of the line-mixing.
These approximations are routinely used in practical treatment of
line-mixing, for example the implementation in the HITRAN database
\cite{10LaTrLaGa}.

\section{Discussion}
The recommendation by the TG of the HTP as the new standard for representing the profile
of high-resolution spectroscopic transitions raises a number of issues which
should be considered. 

First, use of this more complex parameterization to characterize the pressure line shape 
has a consequence not mentioned so far: collisional parameters for gas mixtures are no longer
simple linear combinations of the parameters for the various active molecule-perturber pairs \cite{13NgLiTr}.
This means that, for example, in the terrestrial atmosphere, 
it will be necessary for databases to 
separately specify collisional parameters for perturbations by N$_2$ and O$_2$
rather than just for ``air'' so that separate profiles can be computed and then added.  
As a result of this it will become necessary for
all significant perturbers to be added individually to the databases.

Second, the discussion above has not addressed the temperature
dependence of parameters in the line-profile models considered.
Even for studies of the Earth's atmosphere it is necessary
to consider line profiles in the temperature range of approximately
$200$ to $300$~K. 
Much larger temperature ranges are required for other modelling studies. 
For example, water line profiles are required for the
atmosphere of hot Jupiter exoplanets ($T =$ 1000 -- 1500 K)
\cite{jt521} and probably also for brown dwarfs ($T \leq 3000$ K).
Available experimental studies usually span
ranges of about 100 K to about room temperature,
see Refs. \cite{12BiWa,13CiFoMcLo} for example. 
Obviously, it would be desirable to have experimental studies over
more extended temperature ranges, but these are unlikely to be
forthcoming in the immediate future.  
Conversely, {\it ab initio} spectral shapes calculated by molecular dynamics
simulations \cite{13HaTrNg,13HaSiBo} should be reliable at temperatures above the quantum limit
and can easily be repeated for many temperatures. So
far such studies on water seem to have focused heavily on the room temperature
regime \cite{ntg12,12NgTrGa,13NgTrGaHa,13Mo}.
Extending molecular dynamics simulations to
probe the effects of temperature would clearly be very useful. 
In the absence of experimental determinations, this would allow predictions of the temperature dependence 
of the narrowing and correlation parameters.
 We expect the latter to be reliable, in view of the quality of calculations at room temperature.
Furthermore, the temperature dependences of $\mathit{\Gamma}_2$ and $\mathit{\Delta}_2$, 
like $\mathit{\Gamma}$ and $\mathit{\Delta}$, can be 
investigated by using semi-classical calculations \cite{ntg12,13Mo}.

Third, other tests, both experimental and numerical, would also be useful. 
Issues that should be probed include: 
(a) signal-to-noise limits that the HTP
(and other profiles) are reliable for; (b) tests of extreme heavy -- light
collisions such as water perturbed by He (or H$_2$), such collisions are normally
considered to be ``soft''; (c) tests of extreme
light --- heavy collisions such as water with Xe or SF$_6$;  and (d) tests of
whether collisions with open-shell systems, such as O$_2$, introduce
any new features.

\begin{figure}
\includegraphics[height=130mm]{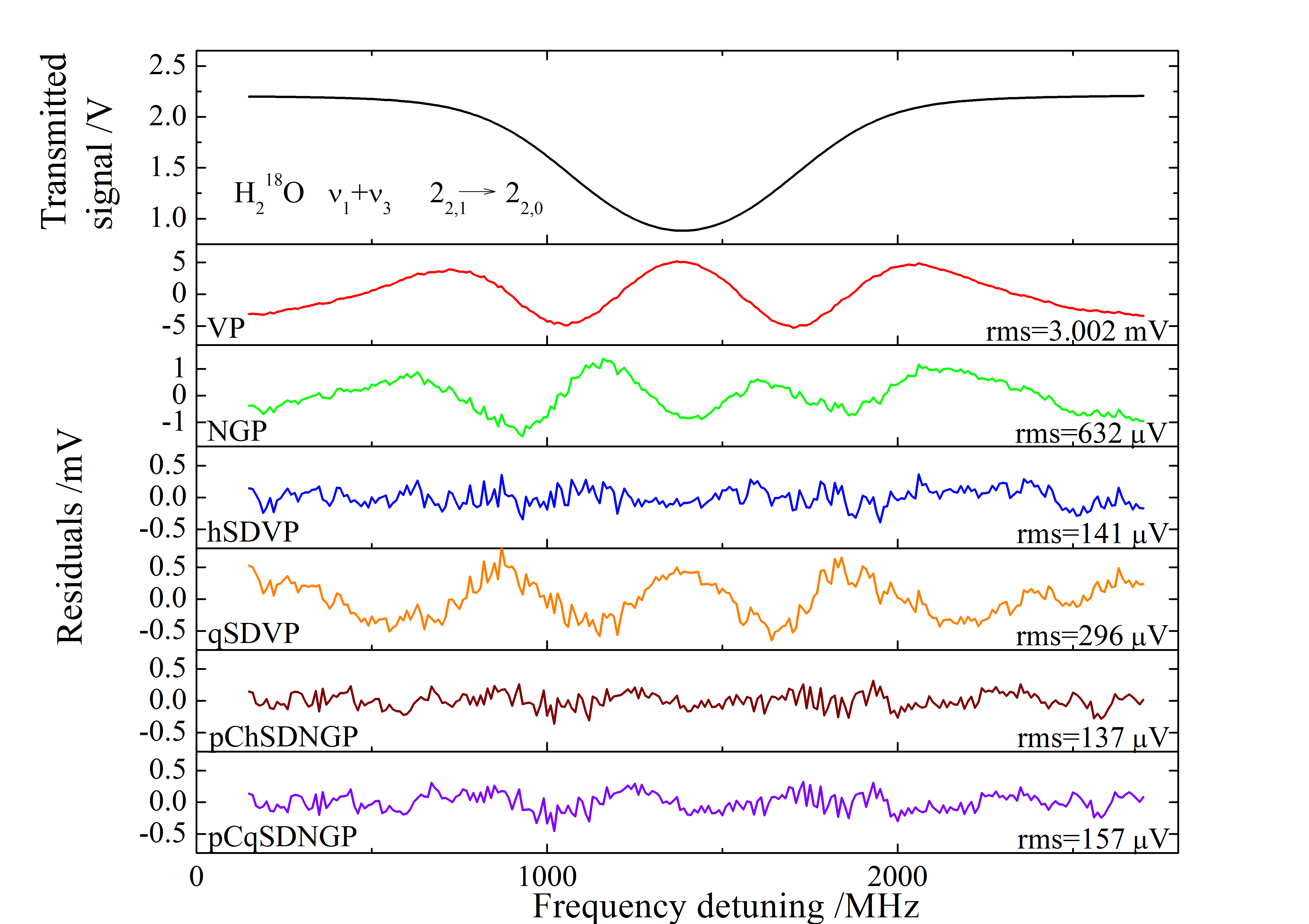}
\caption{Comparison of line-shape fits to the H$_2$$^{18}$O absorption
feature at 7222.298050  cm$^{-1}$  measured at a pressure of 2.70 Torr and 
a temperature of 273.16~K.
Residuals  are given in terms of units of the original
signal: root mean square (rms) values of about 150 $\mu$V  simply reflect
the noise in the original experiment \cite{12DeCaFa}. Note that pCqSDNGP is equivalent to HTP.}
\label{figure:Fig1}
\end{figure}

Finally, the HTP is based on an approximate quadratic treatment of the
speed dependence and of velocity changes. The approximate treatment
of SD can be tested against the more physically-based and more
complicated pChSDNGP, which uses the hypergeometric SD, albeit still
with an approximate treatment of the potential. Such tests have
recently been completed by De Vizia {\it et al.} \cite{14DeFaCa}, see
Fig.~1. They found that the HTP and pChSDNGP fitted their benchmark,
measured, high-precision water line profiles almost equally well.
Furthermore, they found that the retrieved broadening and shifting
parameters were unchanged in the two fits, something not completely
found, for example, when using the SDGP (speed-dependent Galatry
profile) to fit the same data \cite{12DeCaFa}.  However, values for
the parameter $\nu_{\rm VC}$, which can be related to the diffusion
coefficient of the molecule in the perturbing medium, were found to be
physically meaningful for the pChSDNGP fits but not the HTP ones
\cite{14DeFaCa}. Under these circumstances $\nu_{\rm VC}$ must just be
regarded as a useful fit parameter, and nothing more.  From Fig.~1,
one can see that the residual of the fit with the VP varies
between about $\pm 0.5$~\%\ of the peak absorption. These data are for
a specific line and pressure, but much larger effects are observed
\cite{07TrBeDo,12NgTrGa} for other transitions and gas densities.

\section{Final recommendations and conclusions}

The IUPAC Task Group formed by the first 12 authors of this paper makes
the recommendation that the partially Correlated
quadratic-Speed-Dependent Hard-Collision profile \cite{13NgLiTr,13TrNgHa,14TrNgHa,14NgLiTr} should be adopted as
the appropriate line-profile model of high-resolution spectroscopy
moving beyond the VP.
For simplicity we propose calling this the Hartmann--Tran profile (HTP).

The proposed line shape is based on six temperature-dependent,
collisional parameters for each line and perturber plus the Doppler
width, $\mathit{\Gamma}_{\rm D}$, fixed to its theoretical value.  
These parameters give the model the flexibility to include all the major
``non-Voigt'' effects.  
HTP involves parameters with known pressure dependences that can be stored in databases.  
HTP has been demonstrated to lead to an accurate description (0.1 \%\ or better) of
the line shapes for a number of combinations of absorbers and
perturbers \cite{13NgLiTr}, with the exception of H$_2$, which is a
known difficult case.  
Furthermore, HTP can be computed accurately using
only moderate computer time requirement \cite{13TrNgHa}, it is
compatible with current implementations of line mixing, and it can be
initially parameterized using the Voigt parameters already available
in standard data compilations.

Considering the relatively large number of parameters required for the full HTP model and
the correlations between them, fitting measured individual spectra is unlikely to yield a well-constrained
parameter set. 
This means that a multispectrum procedure \cite{95BeRiMaSm} must be used. 
Furthermore, 
to remove the partial correlations between the various parameters, 
it is essential to use spectra recorded in a broad pressure range and with a high signal-to-noise ratio.

There are some open questions about the use of the HTP. 
The most pressing of these concerns the temperature dependence of the
model parameters. 
These have not been the subject of serious 
testing over extended temperature ranges
with temperature-dependent data from either laboratory measurements
or numerical (molecular dynamics and semi-classical) simulations. 
Furthermore, very high signal-to-noise ratio measurements, combined with multispectrum fits of data
recorded at a number of pressures, 
 would help to better define the underlying accuracy of the model.
Finally, tests need to be performed with a variety of collision partners
to determine, for example, the effects of different mass ratios and collisions
with open-shell species such as O$_2$.

One issue not discussed so far is that of the so-called water
continuum \cite{12ShPtRa}.  The water vapour continuum is
characterised by absorption that varies smoothly with wavelength, from
the visible to the microwave.  It is present within the rotational and
vibrational-rotational bands of water vapour, and in the many \lq\lq
windows'' between these bands. The precise relationship between the
water continuum with the far wings of the line profile for individual
transitions and with dimer contributions remains a matter for
discussion.  It is clear that changing the model for the line profile
about the line center has possible consequences on the determination
of the continuum. However, these should be small as local line
contributions are only calculated in a narrow interval before
subtraction from the measured absorption.  Changing the line shape
thus has only a local effect with minor consequences on the broad and
slowly varying continuum.  Furthermore, the main uncertainties in the
continuum determination remain more to do with uncertainties in the
line intensity and broadening coefficients than the line profile.
Nevertheless, for consistency, a continuum should be used in radiative
transfer calculations which has the same local line shape as the one
used in its experimental determination.

Finally, we note that adoption of the HTP will require
significant alteration to the data structures used in databases. Work in this
direction has already started \cite{jt559}. 
Although TG2
was set up explicitly to consider  water, our recommendations and the
modernizations should apply to the line shape used for all molecular species.

\section*{MEMBERSHIP OF SPONSORING BODY}

Membership of the IUPAC Physical and Biophysical Chemistry Division Committee for the period 2014--2015 is as follows:\\
 {\bf President:} R. Marquardt (France);  {\bf Vice President:} A. K. Wilson
(USA);  {\bf Secretary:} A. Friedler (Israel);  {\bf Past President:} K. Yamanouchi
(Japan); {\bf  Titular Members:} K. Bartik (Belgium); A. Goodwin (USA); A. E.
Russell (UK); J. Stohner (Switzerland); Y. H. Taufiq-Yap (Malaysia);
F. van Veggel (Canada);  {\bf Associate Members:} K. Battacharyya (India); A.
G. Cs\'asz\'ar (Hungary); J. de Faria (Portugal); V. Kukushkin (Russia);
\'A. W. Mombr\'u (Uruguay); X. S. Zhao (China/Beijing);  {\bf National
Representatives:} Md. Abu bin Hassan Susan (Bangladesh); H. R. Corti
(Argentina); J. Cejka (Czech Republic); S. Hannongbua (Thailand); S.
J. Kim (Korea); E. Klein (Bulgaria); M. Koper (Netherlands); M.
Korenko (Slovakia); K. E. Laasonen (Sweden); J. E. G. Mdoe (Tanzania);
V. Tomi\v{s}i\'c (Croatia).

\section*{ACKNOWLEDGMENTS}
We thank Dr Linda Brown and Prof Keith Shine for their inputs in the
writing of
this manuscript.  This work was supported by the International Union
of Pure and Applied Chemistry for funding under project 2011-022-2-100
(Intensities and line shapes in high-resolution spectra
of water isotoplogues from    Experiment and Theory). We thank
the Royal Society's Kavli Centre for their hospitality during the
completion of this work.  In addition, this work has received partial
support from
the UK Natural Environment Research Council,
the Royal Society, 
the European Research Council under Advanced Investigator Project 267219,
the Scientific Research Fund of Hungary (grant OTKA NK83583),
the Greenhouse Gas and Climate Sciences Measurements Program of the National Institute of Science and Technology,
the National Science Foundation of the U.S.A. through Grant No.~AGS-1156862,
the Polish National Science Centre, Project no. DEC-2011/01/B/ST2/00491,
the Foundation for Polish Science TEAM Project co-financed by the EU European Regional Development Fund,
the Russian Foundation for Basic Research,
NASA Earth Observing System grant NNX11AF91G and NASA Planetary Atmosphere grant NNX10AB94G,
the NASA laboratory astrophysics program,
the Programme National LEFE (CHAT) of CNRS (INSU),
and the Laboratoire International
Associ\'e SAMIA 
between CNRS (France) and RAS (Russia).

\bibliographystyle{apsrev}

\end{document}